\documentclass[english,floatfix,twocolumn,superscriptaddress,nofootinbib,showpacs,prd]{revtex4-2}
\usepackage[paperwidth=210mm,paperheight=297mm,centering,hmargin=2cm,vmargin=2cm]{geometry}
\usepackage[utf8]{inputenc}
\usepackage[T1]{fontenc}
\usepackage{lmodern}
\usepackage{amsmath}
\usepackage{amssymb}
\usepackage{accents}
\usepackage{graphicx}
\usepackage{subcaption}
\usepackage{esint}
\usepackage{dcolumn}
\usepackage{babel}
\usepackage{xcolor} 
\usepackage{csquotes}
\usepackage{marvosym}
\usepackage{fontawesome5}
\usepackage{hyperref}
\usepackage[overload]{textcase}
\usepackage[symbol,hang,flushmargin]{footmisc}

\makeatletter
\renewcommand*\@fnsymbol[1]{%
  \ifcase#1\or
    \Bat\or
    \Radioactivity\or
    $\star$%
  \else
    \@ctrerr
  \fi
}
\makeatother

\newenvironment{seqn}{\equation\aligned}{\endaligned\endequation}

\newcommand{\be}{\begin{seqn}}
\newcommand{\ee}{\end{seqn}}

\newcommand{\bea}{\begin{eqnarray}}
\newcommand{\eea}{\end{eqnarray}}

\newenvironment{arabicfootnotes}
  {\par\edef\savedfootnotenumber{\number\value{footnote}}
   
   \setcounter{footnote}{0}}
  {\par\setcounter{footnote}{\savedfootnotenumber}}

\begin{document}
%
%
%
%
\title{Spin-2 fields, Lee-Wick Ghosts, and GUP}

\author{Hassan~Alshal}
\email{halshal@scu.edu}
\affiliation{Department of Physics and Engineering Physics, Santa Clara University, Santa Clara, CL 95053, USA}

\author{Elias C. Vagenas}
\email{elias.vagenas@ku.edu.kw}
\affiliation{Department of Physics, College of Science, Kuwait University,\\
Sabah Al Salem University City, P.O. Box 5969, Safat 13060, Shadadiya, Kuwait}

\author{Thomas Van Kortryk}
\email{vankortryk@gmail.com}
\affiliation{Bahamas Advanced Study Institute and Conferences, Long Island, The Bahamas}

\begin{abstract}

\begin{center}
\textbf{ABSTRACT}
\end{center}

\par\noindent
We revisit the structure of higher-derivative spin-2 theories from the perspective of the Generalized Uncertainty Principle (GUP). We show that a minimal GUP deformation of the Fierz-Pauli (FP) action induces a higher-derivative kinetic operator equivalent, at quadratic order, to the spin-2 sector of Stelle's curvature-squared gravity. Via an auxiliary-field formulation, the GUP-generated higher derivative can be recast as a Lee-Wick (LW) partner of the spin-2. We then demonstrate that the same GUP deformation is compatible with the Galileon structure governing the helicity-0 mode in dRGT massive gravity. The GUP corrections reduce to total derivatives, preserving the absence of the Boulware-Deser ghost. Our results unify GUP models, LW quantization, and curvature-squared gravity into a single framework, in which the higher-derivative spin-2 ghost is rendered non-propagating while the nonlinear massive completion remains intact.
\rule{0.82\textwidth}{0.4pt}

In tribute to L. Mezincescu (1946-2025), P. Minkowski (1941-2025), and K. Stelle (1948-2025).
\end{abstract}
\maketitle
\begin{arabicfootnotes}
%
%
%
%

\section{Introduction}
\par\noindent
Stelle's seminal work on the inclusion of quadratic curvature invariants \cite{Stelle:1976gc,Stelle:1977ry} of the terms added to the curvature-squared to the Einstein-Hilbert action
\begin{equation}
  S_{\rm Stelle}
  = \int d^4x\,\sqrt{-g}\left[
      aR
      + b R_{\mu\nu}R^{\mu\nu}
      + c R^2
    \right]
  \label{eq:Stelle-action}
\end{equation}
\par\noindent
renders pure gravity perturbatively renormalizable at the price of introducing
extra massive degrees of freedom. It is famously restores perturbative renormalizability to gravity, but only by enlarging the spectrum with additional massive modes, including a problematic spin-2 excitation. In the linearized
theory around Minkowski space, the spectrum contains, in addition to the
massless spin-2, a massive spin-2 mode with negative residue and a massive scalar. The improved ultraviolet
behavior of the spin-2 propagator at large momenta is entirely due to
these higher-derivative terms \cite{Modesto:2011kw,Tomboulis:2015esa,Asorey:2018wot,Abe:2018rwb,Piva:2023bcf}.
\par\noindent
From the perspective of minimal-length and Generalized Uncertainty
Principle (GUP) scenarios \cite{Amati:1987wq,Gross:1987kza,Amati:1988tn,Konishi:1989wk,Veneziano:1990zh}, it is natural to ask whether such higher-derivative
spin-2 structures can arise as effective descriptions of an underlying
deformation of the phase-space algebra. Quadratic gravity provides a renormalizable, RG-complete gravitational framework in which higher-derivative operators are imposed \cite{Salvio:2018crh}, while GUP-based approaches can be viewed as offering a possible microscopic, kinematical origin for those operators through Planck-scale modifications of momentum space. In GUP models \cite{Kempf:1994su,Kempf:1996nk,Brau:1999uv,Chang:2001bm,Das:2008kaa,Das:2009hs,Myung:2009ur,Pedram:2011gw,Pedram:2012my,Anacleto:2015mma,Das:2010sj,Ali:2009zq,Ali:2010yn,Pedram:2011xj, Nozari:2005ix,Chen:2013tha,Banerjee:2008cf,Moayedi:2010vp, Vagenas:2018pez, Vagenas:2019rai, Vagenas:2019wzd, Vagenas:2020bys, Hemeda:2022dnd}, minimal-length scenarios deform the momentum operator, often introducing curvature in momentum space that manifests as higher-derivative corrections. The quadratic corrections read
\begin{equation}\label{ptop}
  k_\mu \;\to\; k_\mu\bigl(1+\alpha k^2 + \cdots\bigr)
\end{equation}
\par\noindent
which at the level of field theory induce a deformed kinetic operator and propagator with an extra massive pole. This structure is reminiscent of Lee-Wick (LW) theories \cite{Lee:1970iw,Lee:1969fy,Cutkosky:1969fq,Nakanishi:1972wx,Boulware:1983vw,Jansen:1993ji,Jansen:1993jj,Hawking:2001yt,Grinstein:2007mp} and of the
higher--derivative sector of Stelle gravity.

The aim of this work is twofold. First, we show that a GUP-deformed Fierz-Pauli (FP) spin-2 action is equivalent, at quadratic order around flat space, to a particular curvature-squared Lagrangian of Stelle type, i.e., to the spin-2 part of an $R^2$ theory. In this sense, a specific
GUP deformation picks out a special point in the $(b,c)$ parameter
space of eq.\eqref{eq:Stelle-action}, where the linearized spectrum consists of a massless spin-2 plus a massive spin-2 mode only \cite{Fierz:1939ix}. Second, we consider how GUP deformation may affect the massive spin-2 case, particularly the Galileons \cite{Curtright:2020cta} structure of the dRGT generalization of the massless FP spin-2 field \cite{deRham:2010ik}.
\par\noindent
For the first goal, we revisit the quantization of this massive spin-2 mode using LW and PT-symmetric techniques \cite{Bender:1998ke,Bender:1998gh,Bender:1999ek,Bender:2003ve,Bender:2005hf,Mostafazadeh:2003iz,Mostafazadeh:2008pw,Bender:2008vh,Bender:2012ea,Bender:2015uxa,Raidal:2016wop,Bender:2018pbv,Bender:2021fxa,Donoghue:2018lmc, Kuntz:2024rzu}. Rather than adopting an
indefinite-metric state space as in the original Stelle renormalization analysis, we treat the massive spin-2 mode as a LW
partner with complex poles and a deformed integration contour, or
equivalently as a pseudo-Hermitian degree of freedom with respect to a nontrivial metric operator. This provides a GUP-motivated realization of ``LW gravity'' whose quadratic action coincides with Stelle's gravity, but whose ghost is  rendered benign in the LW/PT-symmetric sense.
\par\noindent
Finally, we apply GUP deformation to the dRGT construction \cite{deRham:2010ik,deRham:2010kj,Hinterbichler:2011tt,deRham:2014zqa} to provide a nonlinear, ghost--free massive spin-2 theory, in which the helicity-0 mode $\pi$ is governed in the decoupling limit by Galileon terms. Since massive gravity is a class of Galileon theories \cite{Horndeski:1974wa,Fairlie:2011md,Deffayet:2009wt,Deffayet:2009mn,Curtright:2020cta}, it is worth noting that a wider class of symmetries extends the usual Galilean shift symmetry \cite{Buoninfante:2018lnh}. Instead of standard local derivative Lagrangians, with a finite number of derivatives, nonlocal infinite-derivative actions are provided. For a scalar field, the propagator acquires infinitely many complex-conjugate poles — but, importantly, tree-level unitarity is preserved. The nonlocal generalization preserves the ordinary Galilean shift symmetry. But instead of limiting to finite-derivative self-interactions, it opens up to infinite-derivative operators, which can tame ultraviolet (UV) behavior, e.g. loop divergences, more effectively than local Galileons. Also, quadratic gravity modifies Einstein's theory by adding higher-curvature terms. This gives better ultraviolet behavior, making the gravity theory power-counting renormalizable — but at the cost of introducing a massive spin-2 ghost, from higher-derivative kinetic terms, and typically unitarity problems. Ref. \cite{Buoninfante:2018lnh} aims to retain the good UV behavior, or at least improved behavior, akin to higher-derivative theories — by using infinite-derivative operators — while avoiding ghosts. Indeed, in such nonlocal gravity model, a ghost-free spin-2 propagator is  achieved despite having infinite derivatives. Moreover, and unlike the classical singularities typical in local higher-derivative, or Einstein, gravity, the metric sourced by a point mass is nonsingular, i.e., regular at short distances.
\par\noindent
A priori, the GUP-induced derivative deformation could spoil this structure and reintroduce a Boulware-Deser ghost. We therefore apply the same GUP deformation to the Galileon
building blocks, and show that the $\mathcal{O}(\alpha)$ pieces in the Galileon structure are total derivatives such that the equations of motion for the helicity-0 remain second order at leading order in $\alpha$, and the absence of the Boulware-Deser ghost is preserved in the GUP-deformed decoupling limit. Thus, the GUP deformation considered here reproduces the higher-derivative spin-2 sector of Stelle's $R^2$-gravity, where the massive spin-2 excitation is treated as a LW/ghost partner. Meanwhile, GUP deformation is compatible with the dRGT Galileon structure in the helicity-0
sector, so that a ghost--free nonlinear massive completion remains viable, at least to $\mathcal{O}(\alpha)$.
\par\noindent
For convenience,  we have set $G=\hbar=c=1$ in this paper. 
\section{Explicit GUP $\leftrightarrow$ LW connection}
\par\noindent
In non-anticommutative superspace, SUSY breaking can generate a specific GUP \cite{Faizal:2016zlo} of the form
\begin{equation}
[x_i, p_j] = i\delta_{ij}(1 + \alpha p^2  + \dots)
\end{equation}
\par\noindent
such that the induced higher-derivative terms in the effective fermionic action can be rewritten as a \emph{LW-type extension}; and extra higher-derivative kinetic terms in ordinary Hermitian quantization would correspond to ghost partners.
This means that GUP corrections can be encoded as LW higher-derivative structures, and, conversely, LW models can be viewed as an effective description of some GUP physics. This is the kind of ``GUP inspired by LW'', but it is done for matter fields (fermions) in SUSY, not for gravity itself. It does not yet give a ghost-free spin-2 theory, but the moral is that such mapping
\[
\text{LW partners} \;\longleftrightarrow\; \text{Higher Derivatives} \;\longleftrightarrow\; \text{GUP}
\]
is a consistent one. Inspired by Anselmi--Piva's non-analytic Wick rotation and complex poles \cite{Anselmi:2017lia,Anselmi:2017yux,Anselmi:2018kgz,Anselmi:2018tmf,Anselmi:2018ibi}, we combine GUP with LW quantization to get a plausible strategy for ``GUP-generated LW gravity'', where the extra poles are handled \emph{à la} LW instead of being physical ghosts.
\par\noindent
Several works treat minimal-length or GUP-deformed commutators within PT-symmetric or pseudo-Hermitian frameworks. Ref. \cite{Bagchi:2009wb} shows that minimal-length/GUP-type commutators naturally lead to non-Hermitian operator representations. In addition, it demonstrates how to construct the metric operator that restores unitarity
\begin{equation}\label{nonHermitian}
H^\dagger = \eta H \eta^{-1}~.
\end{equation}
\par\noindent
Meanwhile, Ref. \cite{Dey:2013hda} compares Hermitian and PT-symmetric representations of the same GUP algebra. Moreover, it shows that physical expectation values are representation-independent once the correct metric $\eta$ is chosen. Furthermore, it addresses PT-symmetric noncommutative spaces with minimal area/volume uncertainties, so it constructs PT-symmetric algebras yielding minimal length, minimal area, and even minimal volume scales. Additionally, Refs. \cite{Znojil:2009vn,Znojil:2009th} show how PT-symmetric inner products can introduce a fundamental smearing length in quantum theory.
These works establish that
\begin{equation}\label{deformed}
[x, p] = i\,f(p)
\end{equation}
generically leads to non-Hermitian representations of $x$ and $p$. As GUP operators are typically nonlinear in momentum, their coordinate representation is not Hermitian under the standard inner product. PT-symmetric and pseudo-Hermitian constructions provide a natural arena in which such operators retain real spectra. Thus, there \emph{is} a PT-symmetric/nonstandard quantization branch of GUP. Deformed commutation relations of the form of eq.\eqref{deformed} may lead to momentum operators that are not Hermitian in the usual inner product.  
However, PT-symmetric or pseudo-Hermitian quantum mechanics provides a generalized notion of Hermiticity under which the Hamiltonian remains \emph{quasi-Hermitian} and the spectrum is real. So, we see that minimal-length quantum mechanics can be embedded in a PT-symmetric or pseudo-Hermitian framework. Also, ghost-like negative-norm states can be turned into positive-norm states in the $\eta$-inner product. This reinterpretation does not remove higher derivatives but instead alters the Hilbert-space structure so that unitarity is restored.
\par\noindent
In higher-derivative GUP-induced field theories, extra poles in propagators appear as ghosts under standard Hermitian quantization.  
Nonstandard approaches reinterpret or neutralize these degrees of freedom:
\begin{itemize}
\item 
\textbf{LW approach}:  
Ghost-like poles are made nonphysical by contour prescriptions; they do not appear in asymptotic states.

\item 
\textbf{PT-symmetric/pseudo-Hermitian approach}: 
Wrong-sign kinetic terms can be made positive with respect to a new inner product  
\begin{equation}
\langle \psi | \phi \rangle_\eta = \langle \psi | \eta | \phi \rangle
\end{equation}
eliminating negative norms.

\item 
\textbf{Nonlocality}:  
Using exponential form factors such as $e^{-\alpha \Box}$ can remove extra poles entirely, avoiding ghosts from the beginning \cite{Nouicer:2007jg,Nouicer:2007cw,Kim:2007if}.
\end{itemize}
\par\noindent
Several approaches to GUP can be constructed by drawing inspiration from nonstandard quantization schemes, notably LW theories, PT-symmetric quantum mechanics, and more general pseudo-Hermitian frameworks.  These ideas provide guidance on how one might attempt to control or remove the ghost-like degrees of freedom that frequently arise in higher-derivative or nonlocal gravitational theories.
\par\noindent
LW theories introduce higher-derivative operators whose additional poles appear as massive ``partner'' states with negative norm, but are rendered harmless by a specific contour prescription.  
A GUP whose modified momentum operator contains higher derivatives as in eq.\eqref{ptop} naturally produces propagators of the LW type
\begin{equation}
\frac{1}{k^{2}} \;\longrightarrow\; \frac{1}{k^{2}\,(1 + \alpha k^{2})}
\end{equation}
\par\noindent
with an extra pole at $k^{2} = -1/\alpha$.  
Such a structure is reminiscent of LW ghost partners.  
The central idea is that these extra states may be made nonpropagating or benign if one adopts a LW contour or an equivalent
nonperturbative prescription.
\par\noindent
For a GUP-modified spin-2 theory, one typically obtains kinetic operators containing higher powers of $\Box$, such as
\begin{equation}
\Box\,(1 - \alpha \Box)\,h_{\mu\nu}~.
\end{equation}
\par\noindent
In conventional quantization, this produces ghost poles. However, if the theory is embedded in a pseudo-Hermitian or PT-symmetric framework, one may attempt to define a nontrivial metric operator in the space of gravitational perturbations, shifting the would-be ghost into a non-negative-norm sector.
\par\noindent
Alternatively, one may follow the LW prescription: impose a contour
deformation that avoids the contribution of the ghost pole.  
This can work if:
\begin{enumerate}
\item The ghost pole is complex (LW requirement).
\item It never appears as an asymptotic state.
\item The theory analytic structure allows a consistent deformation.
\end{enumerate}
\section{GUP-inspired higher-derivative Kinetics}
\par\noindent
We consider a real scalar $\phi$ in flat space with signature $(-,+,+,+)$. 
GUP/minimal-length type corrections often give a kinetic operator like
\begin{equation}\label{kinetic}
    K(\Box) = \Box (1-\ell_{\text{min}}^2 \Box)
\end{equation}
\par\noindent
where $\ell_{\text{min}}$ is some minimal length of the order of the Planck length $\ell_{\text{Pl}}$.
\par\noindent
Writing the quadratic action as
\begin{equation}
S \;=\; \frac12 \int d^4x\; \phi\, K(\Box)\,\phi
= \frac12\int d^4x\; \phi\,\Box(1-\ell_{\text{min}}^2\Box)\phi
\end{equation}
\par\noindent
and integrating by parts, we  obtain
\begin{equation}
S = -\frac12\int d^4x\; (\partial_\mu\phi)(\partial^\mu\phi)
- \frac{\ell_{\text{min}}^2}{2}\int d^4x\; (\Box\phi)^2~.
\end{equation}
\par\noindent
Now we visit the momentum space and, thus, $\Box \to -k^2$. 
The inverse propagator is written as
\begin{equation}
K(k^2) = -k^2 \big(1 + \ell_{\text{min}}^2 k^2\big)
\end{equation}
\par\noindent
so the propagator reads
\begin{align}
\Delta(k) &= \frac{i}{K(k^2)+i\epsilon}
= \frac{i}{-k^2(1+\ell_{\text{min}}^2 k^2)+i\epsilon}\notag\\
&= -\,\frac{i}{k^2(1+\ell_{\text{min}}^2 k^2) - i\epsilon}~.
\end{align}
\par\noindent
Ignoring the overall minus as it just flips the sign convention for the whole action, the pole structure is what matters and, thus, we have
\begin{equation}
\frac{1}{k^2(1+\ell_{\text{min}}^2 k^2)}
= \frac{A}{k^2} + \frac{B}{k^2 + 1/\ell_{\text{min}}^2}~.
\end{equation}
\par\noindent
We solve for the coefficients to get
\begin{equation}
1 = A(k^2+1/\ell_{\text{min}}^2) + B k^2
\;\Rightarrow\;
\begin{cases}
A + B = 0,\\[2pt]
\dfrac{A}{\ell_{\text{min}}^2} = 1,
\end{cases}\notag
\end{equation}
\begin{equation}
A = +\ell_{\text{min}}^2,\; B = -\ell_{\text{min}}^{2}~.
\end{equation}
\par\noindent
So, the propagator becomes
\begin{align}
\Delta(k) \propto \ell_{\text{min}}^2 \left( \frac{1}{k^2} - \frac{1}{k^2 + 1/\ell_{\text{min}}^2}\right)
\end{align}
up to an overall positive factor. This is
\begin{equation}
\Delta(k) \sim \frac{1}{k^2} \;-\; \frac{1}{k^2 + M^2},
\qquad M^2 \equiv 1/\ell_{\text{min}}^{2}~.
\end{equation}
\par\noindent
So, there is a massless pole at $k^2=0$ with zero residue. Meanwhile, the massive pole at $k^2=-M^2$ with negative residue yields a ghost in standard Hermitian quantization for the mostly positive metric signature.
\section{From the $K(\Box)$ to GUP}
\par\noindent
We need to relate the $f(p)$ in eq.\eqref{deformed} to the kinetic operator of eq.\eqref{kinetic}. According to Ref. \cite{Abdelkhalek:2016nyn}, one can consider a scalar setup with canonical variables $(x,k)$ obeying
\begin{equation}
[x,k] = i~.
\end{equation}
\par\noindent
Then, a deformed physical momentum \(p = p(k)\) is obtained. Thus, the GUP commutator becomes as in eq.\eqref{deformed} so that
\begin{equation}\label{dp/dk}
f(p) = \frac{dp}{dk}~.
\end{equation}
\par\noindent
In momentum space, eq.\eqref{kinetic} corresponds to the eigenvalue
\begin{equation}
K(k^{2}) = k^{2}\bigl(1 + \ell^2_{\text{min}} k^{2}\bigr)~.
\end{equation}
\par\noindent
So, we identify the momentum as
\begin{equation}\label{p(k)}
p(k) = k\sqrt{1 + \ell^2_{\text{min}} k^{2}}
\end{equation}
\par\noindent
and eq.\eqref{dp/dk} becomes
\begin{equation}
\frac{dp}{dk}
  = \sqrt{1 + \ell^2_{\text{min}} k^{2}}
    + \frac{\ell^2_{\text{min}} k^{2}}{\sqrt{1 + \ell^2_{\text{min}} k^{2}}}
  = \frac{1 + 2\ell^2_{\text{min}} k^{2}}{\sqrt{1 + \ell^2_{\text{min}} k^{2}}}~.
\end{equation}
\par\noindent
Thus, the GUP commutator is 
\begin{equation}\label{f(p(k))}
f\bigl(p(k)\bigr)
  = \frac{1 + 2\ell^2_{\text{min}} k^{2}}{\sqrt{1 + \ell^2_{\text{min}} k^{2}}}~.
\end{equation}
\par\noindent
Now, we square eq.\eqref{p(k)} and solve for $k(p)$ such that eq.\eqref{f(p(k))} becomes
\begin{equation}
f(p)
  = \frac{\sqrt{1 + 4\ell^2_{\text{min}} p^{2}}}
         {\sqrt{ \frac{1 + \sqrt{1 + 4\ell^2_{\text{min}} p^{2}}}{2} }}~.
\end{equation}
\par\noindent
We expand by setting \(z = \ell^2_{\text{min}} p^{2}\) to get
\begin{align}
f(p)
  &= \frac{1 + 2z + \mathcal{O}(z^{2})}{1 + \frac{1}{2}z + \mathcal{O}(z^{2})}
  = 1 + \frac{3}{2} z + \mathcal{O}(z^{2})\notag\\
 &= 1 + \frac{3}{2}\,\ell^2_{\text{min}} p^{2} + \mathcal{O}(\ell^4_{\text{min}}p^{4})
\end{align}
\par\noindent
and, to leading order, the GUP algebra becomes
\begin{equation}\label{GUPalgebra}
[x,p] = i\left(1 + \frac{3}{2}\,\ell^2_{\text{min}} p^{2} + \cdots\right)~.
\end{equation}
\par\noindent
Therefore, we can define a generalized uncertainty principle of the form
\begin{equation}
    \Delta x\Delta p \geq \frac{1}{2} (1+\frac{3}{2}(\ell_{\text{min}}p)^2)
\end{equation}
\par\noindent
with the optimal momentum spread to be
\begin{equation}\label{opt.p}
\displaystyle\Delta p_{\text{opt}}=\sqrt{\frac{2}{3}}\frac{1}{\ell_{\text{min}}}~.
\end{equation}
\section{GUP-induced higher-derivative kinetic operator for spin-2 fields}
\par\noindent
In this section, we generalize the scalar GUP construction to a massless spin-2 field, starting from the FP action and engineering a GUP-induced higher-derivative kinetic operator acting on the spin-2 sector.  We consider the linearized FP Lagrangian density for the massless case
\begin{align}\label{FP}
\mathcal{L}_{\mathrm{FP}}
  =& \ \frac{1}{4}\,\partial_\lambda h_{\mu\nu}\,\partial^\lambda h^{\mu\nu}
    - \frac{1}{2}\,\partial_\mu h^{\mu\rho}\,\partial^\nu h_{\nu\rho}\notag\\
    &+ \frac{1}{2}\,\partial_\mu h\,\partial_\nu h^{\mu\nu}
    - \frac{1}{4}\,\partial_\lambda h\,\partial^\lambda h
\end{align}
\par\noindent
where $h_{\mu\nu}$ denotes the small fluctuations around Minkowski spacetime as $g_{\mu\nu} = \eta_{\mu\nu} + \,h_{\mu\nu}$, and $h \equiv \eta^{\mu\nu}h_{\mu\nu}$. We now introduce the trace-reversed field
\begin{equation}
\label{eq:barh}
\bar{h}_{\mu\nu} \;\equiv\; h_{\mu\nu} - \frac{1}{2}\,\eta_{\mu\nu} h
\end{equation}
and impose the de Donder (harmonic) gauge condition
\begin{equation}
\label{eq:dedonder}
\partial^\mu \bar{h}_{\mu\nu} \;=\; 0~.
\end{equation}
\par\noindent
In this gauge, the equations of motion reduce to
\begin{equation}
\label{eq:FP_eom}
\Box \bar{h}_{\mu\nu} \;=\; 0
\end{equation}
\par\noindent
and the quadratic action can be written schematically as
\begin{equation}
\label{eq:FP_quad}
S_{\mathrm{FP}} \;=\; \frac{1}{2} \int d^4x\;
h_{\mu\nu}\,\mathcal{E}^{\mu\nu\rho\sigma} h_{\rho\sigma},
\end{equation}
\par\noindent
where $\mathcal{E}^{\mu\nu\rho\sigma}$ is the Lichenrowicz operator.
In the de Donder gauge and momentum space, one has
\begin{equation}
\label{eq:E_p}
\mathcal{E}^{\mu\nu\rho\sigma}(k)
 = -k^2\,P^{\,\mu\nu\rho\sigma}
\end{equation}
\par\noindent
where $P^{\,\mu\nu\rho\sigma}$ is the standard spin-2 projector
\begin{equation}
P^{\mu\nu\rho\sigma}=\eta^{\mu\rho}\eta^{\nu\sigma}+\eta^{\mu\sigma}\eta^{\nu\rho}-\eta^{\mu\nu}\eta^{\rho\sigma}~.
\end{equation}
\par\noindent
Thus, the FP kinetic operator in the spin-2 sector is
\begin{equation}
\label{eq:K_FP}
K^{\mu\nu\rho\sigma}_0(k^2) \;=\; -k^2\,P^{\,\mu\nu\rho\sigma}~.
\end{equation}
\par\noindent
In momentum space, the deformed spin-2 kinetic operator is obtained by replacing $k^2$ in \eqref{eq:K_FP} with $p^2(k)$
\begin{align}
\label{eq:K_FP_deformed_p}
K^{\mu\nu\rho\sigma}(k^2) 
  =-\,p^2(k)\,P^{\,\mu\nu\rho\sigma}
  =-\,k^2\bigl(1 + \ell^2_{\text{min}} k^2\bigr) P^{\,\mu\nu\rho\sigma}~.
\end{align}
\par\noindent
Going back to position space, with $k^2 \rightarrow -\Box$, we obtain the GUP-induced higher-derivative kinetic operator
\begin{equation}
\label{eq:K_FP_deformed_Box}
K^{\mu\nu\rho\sigma}(\Box)
  \;=\; \Box\,(1 - \ell^2_{\text{min}} \Box)\,P^{\,\mu\nu\rho\sigma}~.
\end{equation}
\par\noindent
Therefore, the GUP-modified FP action in the spin-2 sector can be written as
\begin{equation}
\label{eq:S_FP_deformed}
S_{\mathrm{FP}}^{(\ell^2_{\text{min}})}
  \;=\; \frac{1}{2}\int d^4x\;
   \bar{h}_{\mu\nu}\, \Box(1 - \ell^2_{\text{min}}\Box)\,P^{\,\mu\nu\rho\sigma}\,\bar{h}_{\rho\sigma}~.
\end{equation}
\par\noindent
This corresponds to one massless spin-2 mode with positive residue and one massive spin-2 mode with opposite sign residue, which in standard Hermitian quantization would be interpreted as a spin-2 ghost. In LW or PT-symmetric quantization scheme, the massive pole can be treated as a LW partner rather than a propagating ghost in the  asymptotic spectrum.
\par\noindent
According to Refs. \cite{Todorinov:2018arx,Bosso:2020fos,Bosso:2020jay,Bosso:2024nmn}, the momentum can be deformed into
\begin{equation}
    p_{\mu}\to p_{\mu}=k_{\mu}(1+\alpha k^2)
\end{equation}
\par\noindent
which matches with the GUP algebra eq.\eqref{GUPalgebra}. In order to use that definition, we set $\displaystyle\alpha=\frac{3}{2}\ell^2_{\text{min}}$. Then, we deform the differential operator
\begin{equation}\label{deformedPartial}
    \partial_{\mu}\to \partial_{\mu}(1-\alpha\Box)
\end{equation}
\par\noindent
such that
\begin{equation}
    \Box\to \Box-2\alpha\Box^2+\mathcal{O}(\alpha^2)\sim\Box(1-2\alpha\Box)~.
\end{equation}
\par\noindent
To make the last result similar to eq.\eqref{kinetic}, one can define a new GUP factor $\beta \equiv 2\alpha$. But we can continue with the last result as it is. Now in the light of the last deformation, eq.\eqref{eq:FP_quad} and eq.\eqref{eq:FP_eom} suggest that the Lichenrowicz operator should be deformed too as
\begin{equation}
    \mathcal{E}^{\mu\nu\rho\sigma}\to(1-2\alpha\Box)\mathcal{E}^{\mu\nu\rho\sigma}
\end{equation}
\par\noindent
such that
\begin{equation}
    \mathcal{L}^{\alpha}=\mathcal{L}_{\text{FP}}-2\alpha h_{\mu\nu}\mathcal{E}^{\mu\nu\rho\sigma}h_{\rho\sigma}~.
\end{equation}
\par\noindent
The gauge condition should also be deformed into
\begin{equation}
    \partial_{\mu}\bar{h}_{\mu\nu}=0\to\partial_{\mu}(1-\alpha\Box)\bar{h}_{\mu\nu}=0~.
\end{equation}
\par\noindent
But since $(1-\alpha\Box)$ is invertible, then the deformation should neither affect the gauge nor the degrees of freedom count.
\par\noindent
Consequently, the propagator is now
\begin{equation}
    D^{(\alpha)}_{\mu\nu\rho\sigma}=\frac{iP_{\mu\nu\rho\sigma}}{k^2(1+2\alpha k^2)+i\epsilon}=iP_{\mu\nu\rho\sigma}\bigg[\frac{1}{k^2}-\frac{1}{\frac{1}{2\alpha}+k^2}\bigg]
\end{equation}
\par\noindent
with a usual massless pole at $k^2=0$ and LW pole at $\displaystyle k^2=-1/2\alpha=-\frac{1}{3\ell^2_{\text{min}}}$, or%
\begin{equation}
\displaystyle |k_{\text{LW}}|=\sqrt{\frac{1}{3\ell^2_{\text{min}}}}~.
\end{equation}
\par\noindent
A comparison between the above result and the one in eq.\eqref{opt.p} reveals that 
\begin{equation}
\Delta p_{\text{opt}}\sim|k_{\text{LW}}|
\end{equation}
\par\noindent
which is expected as the LW pole corresponds to a characteristic minimal length/Planck-ish length scale. The corresponding momentum scale is of the same order as, but distinct from, the optimal momentum uncertainty derived from the GUP.
\section{Explicit 4-derivative correction to spin-2}
\par\noindent
Now we recall the linearized Ricci tensor
\begin{equation}
R_{\mu\nu}
=
\frac{1}{2}
\left(
\partial_{\rho}\partial_{\mu}h^{\rho}{}_{\nu}
+
\partial_{\rho}\partial_{\nu}h^{\rho}{}_{\mu}
-
\partial_{\mu}\partial_{\nu}h
-
\Box h_{\mu\nu}
\right)
\end{equation}
\par\noindent
and the Ricci scalar
\begin{equation}
R
=
\partial_{\mu}\partial_{\nu}h^{\mu\nu}
-
\Box h~.
\end{equation}
\par\noindent
Also, the Einstein tensor reads
\begin{equation}
G_{\mu\nu}
=
R_{\mu\nu}
-\frac{1}{2}\eta_{\mu\nu}R
=
-\frac{1}{2}\mathcal{E}_{\mu\nu}{}^{\rho\sigma}h_{\rho\sigma}~.
\end{equation}
\par\noindent
Then, the 4-derivative term can be written as
\begin{equation}
\mathcal{L}^{\Box^2}
=
-\alpha
h_{\mu\nu}\Box \mathcal{E}^{\mu\nu\rho\sigma}h_{\rho\sigma}
=
+2\alpha
h_{\mu\nu}\,\Box\,G^{\mu\nu}~.
\end{equation}
\par\noindent
Integrating by parts twice using self-adjointness of \( \Box \) and of the linearized Einstein operator, the above 4-derivative term becomes a curvature-squared Lagrangian
\begin{equation}
\mathcal{L}^{\Box^2}
\sim
\alpha
\left(
R_{\mu\nu}R^{\mu\nu}
-
\frac{1}{3}R^{2}
\right)~.
\end{equation}
\par\noindent
The $\mathcal{L}^{\Box^2}$ terms are the spin-2 part of the usual quadratic curvature invariants as the part generating the $1/k^{4}$ propagator. So an explicit “FP + GUP” quadratic Lagrangian is
\begin{equation}
\mathcal{L}^{\alpha}
\sim
\mathcal{L}_{\text{FP}}
+
\ell_{\text{min}}^{2}
\left(
R_{\mu\nu}R^{\mu\nu}
-
\frac{1}{3}R^{2}
\right)~.
\end{equation}
\section{Auxiliary Field Diagonalization}
\par\noindent
We just tailored the higher-derivative FP kinetic term as
\begin{equation}
\mathcal{L}^{\alpha}
= 
\frac{1}{2}\, h\,\mathcal{E}\,h
+ 
\frac{1}{2}4\alpha(\mathcal{E}h)_{\mu\nu}(\mathcal{E}h)^{\mu\nu}
\end{equation}
\par\noindent
where $\mathcal{E}_{\mu\nu}{}^{\rho\sigma}$ is the linearized Einstein operator and contractions and integrals are implicit. Then, we introduce an auxiliary symmetric tensor $\psi_{\mu\nu}$ and take the \textit{local} two-field Lagrangian
\begin{equation}
\mathcal{L}_{\text{aux}}
=
\frac{1}{2}\,h_{\mu\nu}\,\mathcal{E}^{\mu\nu\rho\sigma}h_{\rho\sigma}
-
\frac{1}{2}\,\psi_{\mu\nu}\psi^{\mu\nu}
+
\sqrt{2\alpha}\,\psi_{\mu\nu}\,\mathcal{E}^{\mu\nu\rho\sigma}h_{\rho\sigma}~.
\end{equation}
\par\noindent
Then, varying  with respect to \ $\psi$, we obtain
\begin{equation}
\psi_{\mu\nu} = \sqrt{2\alpha}\,(\mathcal{E}h)_{\mu\nu}
\end{equation}
\par\noindent
and, plugging back, $\mathcal{L}^{\alpha}$ is reproduced.
In addition, based on the gauge symmetry, the variation of the fields reads
\begin{equation}
\delta h_{\mu\nu} = \partial_\mu \xi_\nu + \partial_\nu \xi_\mu,
\qquad
\delta \psi_{\mu\nu} = 0.
\end{equation}
\par\noindent
We also introduce an auxiliary field $\chi$ to eliminate the higher derivative, or some equivalent linear combination, so the kinetic terms with a linear field redefinition is diagonalized. So, in this basis,  $\Phi$ is a healthy massless scalar while $\Psi$ is a massive scalar with wrong-sign kinetic term, which results in LW/ghost partner. Now, we define the new fields
\begin{equation}
\Phi_{\mu\nu} \equiv h_{\mu\nu} + \sqrt{2\alpha}\,\psi_{\mu\nu},
\qquad
\Psi_{\mu\nu} \equiv \psi_{\mu\nu~}
\end{equation}
\par\noindent
and utilize the self-adjointness of \(\mathcal{E}\) such that 
\(\Phi \mathcal{E} \Psi = \Psi \mathcal{E} \Phi\). Then, the Lagrangian of the auxiliary field is 
\begin{equation}
\begin{aligned}
\mathcal{L}_{\text{aux}}
&= \frac{1}{2} (\Phi - \sqrt{2\alpha} \Psi)\,\mathcal{E}\,(\Phi - \sqrt{2\alpha}\Psi)\\
   & \quad - \frac{1}{2}\,\Psi^2
   + \sqrt{2\alpha}\,\Psi\,\mathcal{E}\,(\Phi - \sqrt{2\alpha}\Psi) \\[4pt]
&= \frac{1}{2}\,\Phi\mathcal{E}\Phi
   - \sqrt{2\alpha}\,\Phi\mathcal{E}\Psi
   + \alpha\,\Psi\mathcal{E}\Psi\\
   & \quad - \frac{1}{2}\,\Psi^2
   + \sqrt{2\alpha}\,\Psi\mathcal{E}\Phi
   - 2\alpha\,\Psi\mathcal{E}\Psi~.
\end{aligned}
\end{equation}
\par\noindent
It is evident that he mixed terms cancel out
\begin{equation}
-\sqrt{2\alpha}\,\Phi\mathcal{E}\Psi 
\;+\;
\sqrt{2\alpha}\,\Psi\mathcal{E}\Phi 
= 0
\end{equation}
while the \(\Psi\Psi\) pieces can be combined as
\begin{equation}
\alpha\Psi\mathcal{E}\Psi - 2\alpha\,\Psi\mathcal{E}\Psi
= -\alpha\Psi\mathcal{E}\Psi~.
\end{equation}
\par\noindent
Therefore, the Lagrangian of the auxiliary field now reads
\begin{equation}
\mathcal{L}_{\text{aux}}
= \frac{1}{2}\,\Phi_{\mu\nu}\,\mathcal{E}^{\mu\nu\rho\sigma}\Phi_{\rho\sigma}
- \alpha\,\Psi_{\mu\nu}\,\mathcal{E}^{\mu\nu\rho\sigma}\Psi_{\rho\sigma}
- \frac{1}{2}\,\Psi_{\mu\nu}\Psi^{\mu\nu}~.
\end{equation}
\par\noindent
It should be pointed out that, on the TT spin-2 sector, \(\mathcal{E} \to -\Box\), so this becomes schematically
\begin{equation}
\mathcal{L}^{\text{TT}}
= -\frac{1}{2}\,\Phi\,\Box\,\Phi
  + \alpha\,\Psi\,\Box\,\Psi
  - \frac12\,\Psi^{2}~.
\end{equation}
\par\noindent
If we do another field redefinition
\begin{equation}
\Psi_{\mu\nu} \;\to\; \frac{1}{\sqrt{2\alpha}}\,\Psi_{\mu\nu}~,
\end{equation}
\par\noindent
then, the Lagrangian of the TT sector becomes
\begin{equation}
\mathcal{L}^{\text{TT}}
= -\frac{1}{2}\,\Phi\,\Box\,\Phi
  + \frac{1}{2}\,\Psi\,\Box\,\Psi
  - \frac{1}{4\alpha}\,\Psi^{2}~,
\end{equation}
or
\begin{equation}
\mathcal{L}^{\text{TT}}
= -\frac12\,\Phi_{\mu\nu}\Box\Phi^{\mu\nu}
  + \frac12\,\Psi_{\mu\nu}\bigl(\Box - M^2\bigr)\Psi^{\mu\nu}
\end{equation}
\par\noindent
where $M^2=1/2\alpha$, 
$\Phi_{\mu\nu}$ is massless spin-2 with right-sign kinetic term, and $\Psi_{\mu\nu}$ is massive spin-2 with wrong-sign kinetic term as LW / ghost partner.
We couple the original $h_{\mu\nu}$ to matter in the usual way
\begin{equation}
\mathcal{L}_{\text{int}}
= \frac{\kappa}{2}\,h_{\mu\nu}\,T^{\mu\nu}
\end{equation}
\par\noindent
where $T^{\mu\nu}$ is the energy-momentum tensor and $\kappa$ is the coupling constant. Then, in terms of $\Phi,\Psi$ we have
\begin{equation}
\mathcal{L}_{\text{int}}
= \frac{\kappa}{2}\bigl(\Phi_{\mu\nu} - \sqrt{2\alpha}\,\Psi_{\mu\nu}\bigr)T^{\mu\nu}~.
\end{equation}
\par\noindent
So, the healthy spin-2 $\Phi_{\mu\nu}$ couples to $T^{\mu\nu}$ while the LW spin-2 $\Psi_{\mu\nu}$ couples with a suppressed, opposite-sign coupling  $-(\kappa\sqrt{2\alpha}/2)$.

\par\noindent
In the static, weak-field limit, this leads to a Newtonian potential of the form
\begin{equation}
V(r) \sim -\frac{G m_1 m_2}{r}\left(1 - e^{-M r}\right),
\qquad
M = \frac{1}{\sqrt{2\alpha}}
\end{equation}
\par\noindent
which is the usual $1/r$ attraction plus a repulsive Yukawa-type correction from $\Psi$. Under LW quantization, $\Psi$ never appears as an external particle, but it does modify the internal propagator and hence the short-distance gravitational potential.
\section{LW quantization: how to tame the ghost}
\par\noindent
In standard Hermitian quantization, $\Psi$ has negative norm or negative energy, which gives an unstable vacuum and loss of unitarity. The LW idea is as follows:
\begin{enumerate}

\item Give the ghost a complex mass by deforming the parameters slightly
   \begin{equation}
   M^2 \;\to\; M^2 - i\Gamma M
   \end{equation}
   so the pair of ghost poles move off the real axis
   \begin{equation}
   p^2 = M^2 - i\Gamma M \quad\text{and}\quad p^2 = M^2 + i\Gamma M~.
   \end{equation}
But microcausality is violated only at scales \( \sim \ell_{\text{min}} \), because the ghost contribution introduces exponentially suppressed acausal terms
\begin{equation}
[ h_{\mu\nu}(x) , h_{\rho\sigma}(y) ] \neq 0
\quad
\text{as}
\quad
(x-y)^{2}<0
\ \text{and}\ |x-y|\lesssim \ell_{\text{min}}.
\end{equation}
At macroscopic distances, this violation becomes negligible.
\item Define the LW contour $C_{\rm LW}$ for loop integrals so that ordinary poles from $\Phi$ are treated with a usual Feynman prescription, and complex conjugate ghost poles are bypassed in such a way that no negative-norm states appear in the asymptotic Hilbert space. Also, the amplitudes remain real and satisfy the optical theorem (perturbative unitarity).

\item In coordinate space, the LW quantization induces microscopic violations of standard microcausality at distances around $ 1/M\sim \ell_{\text{min}}$, but macroscopic causality and unitarity are preserved.

\end{enumerate}
\par\noindent
So classically, the LW partner looks like a ghost. But quantum-mechanically, with LW contour, that mode never appears as external state; rather it acts more like Pauli-Villars regulator dynamically generated by the GUP. In other words, we keep higher-derivative UV softening, but remove physical ghost states.
\par\noindent
Alternatively, instead of changing the contour, we can reinterpret the wrong-sign sector as part of a non-Hermitian but PT-symmetric Hamiltonian. In the $(\Phi,\Psi)$ basis, the Hamiltonian is not positive-definite with the usual inner product. But we can look for a metric operator $\eta$ as in eq.\eqref{nonHermitian} such that $H$ is $\eta$-pseudo-Hermitian. Then, the inner product $\langle\cdot|\eta|\cdot\rangle$ restores positivity of norms and unitarity of time evolution, at the price of a nontrivial Hilbert-space metric - exactly what people do for minimal-length/GUP Hamiltonians in PT-symmetric QM. So, we can think of the GUP-induced LW partner $\Psi$ as a hidden PT-symmetric degree of freedom: ghost-like in the naive metric, but not in the physical one.

But for gravity, the story is technically harder but conceptually parallel as moving from GUP to higher-derivative spin-2 action with extra poles requires rewriting them in multi-field form, then quantize the extra poles à la LW or PT-symmetric methods.
\section{\MakeLowercase{d}RGT Massive spin-2 and GUP}
\par\noindent
In the dRGT resummations of massive gravity, the spin-2 field $h_{\mu\nu}$ is covariantized into
\begin{equation}
H_{\mu\nu}=h_{\mu\nu}+\partial_{\mu}\pi_{\nu}+\partial_{\nu}\pi_{\mu}-\eta^{\alpha\beta}\partial_{\mu}\pi_{\alpha}\partial_{\nu}\pi_{\beta}
\end{equation}
or
\begin{equation}
H_{\mu\nu}=h_{\mu\nu}+2\Pi_{\mu\nu}-\eta^{\alpha\beta}\Pi_{\mu\alpha}\Pi_{\nu\beta}
\end{equation}
\par\noindent
using the $U(1)$ field $\pi_{\mu}=\partial_{\mu}\pi$, and with $\Pi_{\mu\nu}=\partial_{\mu}\partial_{\nu}\pi$.
Then, the deformed second derivative, to $\mathcal{O}(\alpha)$, in the Galileon structure is defined as
\begin{equation}\label{Galileon2nd}
    \accentset{(\alpha)}{\mathcal{L}}_{(2)}=[\accentset{(\alpha)}{\Pi}]^2-[\accentset{(\alpha)}{\Pi}^2]
\end{equation}
\par\noindent
where after applying eq.\eqref{deformedPartial} on the $\Pi$, we can define the deformed $\Pi$ structure as
\begin{equation}\label{Pi}
\accentset{(\alpha)}{\Pi}\equiv\eta^{\mu\nu}\accentset{(\alpha)}{\Pi}_{\mu\nu}=\Pi-2\alpha\Box\Pi+\mathcal{O}(\alpha^2)~.
\end{equation}
\par\noindent
Next, we square the above result to obtain
\begin{equation}\label{Pi^2}
    [\accentset{(\alpha)}{\Pi}]^2=[\Pi]^2-4\alpha\Pi\Box\Pi=[\Pi]^2-4\alpha\Box\pi\Box^2\pi
\end{equation}
\par\noindent
while the second term in the $\accentset{(\alpha)}{\mathcal{L}}_{(2)}$ reads
\begin{equation}
    [\accentset{(\alpha)}{\Pi}^2]=[\Pi^{2}]-4\alpha\partial_{\mu}\partial^{\alpha}\pi\Box\partial_{\alpha}\partial^{\mu}\pi~.
\end{equation}
\par\noindent
By substituting the last two results into eq.\eqref{Galileon2nd}, we obtain
\begin{equation}\label{Galileon2ndAlpha}
    \accentset{(\alpha)}{\mathcal{L}}_{(2)}=[\Pi]^2-[\Pi^{2}]-4\alpha\big(\Box\pi\Box^2\pi-\partial_{\mu}\partial^{\alpha}\pi\Box\partial_{\alpha}\partial^{\mu}\pi\big)~.
\end{equation}
\par\noindent
The $\alpha$ term in the last equation becomes a total derivative upon being integrated by part. Similarly, for the GUP-deformed third derivative, it is defined as
\begin{equation}\label{Galileon3rd}
    \accentset{(\alpha)}{\mathcal{L}}_{(3)}=[\accentset{(\alpha)}{\Pi}]^3-3[\accentset{(\alpha)}{\Pi}][\accentset{(\alpha)}{\Pi}^{2}]+2[\accentset{(\alpha)}{\Pi}^{3}]~.
\end{equation}
\par\noindent
The first term in eq.\eqref{Galileon3rd} can be expanded as
\begin{equation}
    [\accentset{(\alpha)}{\Pi}]^3=[\Pi]^3-6\alpha(\Box\pi)^2\Box^2\pi~.
\end{equation}
\par\noindent
The second term in eq.\eqref{Galileon3rd} is already provided in eq.\eqref{Pi} and eq.\eqref{Pi^2}, while the third term in eq.\eqref{Galileon3rd} can be expanded as
\begin{equation}
    2[\accentset{(\alpha)}{\Pi}^{3}]=2\big([\Pi^{3}]-6\alpha\partial_{\mu}\partial_{\alpha}\pi\partial^{\alpha}\partial^{\rho}\pi\Box\partial_{\rho}\partial^{\mu}\pi\big)~.
\end{equation}
\par\noindent
Then, we substitute these results in eq.\eqref{Galileon3rd} to get
\begin{align}\label{Galileon3rdAlpha}
    \accentset{(\alpha)}{\mathcal{L}}_{(3)}&=[\Pi]^3-3[\Pi][\Pi^{2}]+2[\Pi^{3}]\notag\\
    & \quad -6\alpha(\Box\pi)^2\Box^2\pi\notag\\
    & \quad-3\alpha\big(\Box\pi\partial_{\mu}\partial^{\nu}\pi\Box\partial^{\mu}\partial_{\nu}\pi+2(\Box\pi)^2\Box^2\pi\big)\notag\\
    & \quad-12\alpha\partial_{\mu}\partial_{\alpha}\pi\partial^{\alpha}\partial^{\rho}\pi\Box\partial_{\rho}\partial^{\mu}\pi
\end{align}
\par\noindent
and, like the $\alpha$ terms in eq.\eqref{Galileon2ndAlpha}, we can treat those $\alpha$ terms in eq. \eqref{Galileon3rdAlpha} as total derivative terms. 
\par\noindent
The same can be expected for the structure of $\accentset{(\alpha)}{\mathcal{L}}_{(4)}$
\begin{align}\label{Galileon4rd}
    \accentset{(\alpha)}{\mathcal{L}}_{(4)}&=[\accentset{(\alpha)}{\Pi}]^4-6[\accentset{(\alpha)}{\Pi}]^2[\accentset{(\alpha)}{\Pi}^{2}]+8[\accentset{(\alpha)}{\Pi}][\accentset{(\alpha)}{\Pi}^{3}]+3[\accentset{(\alpha)}{\Pi}^{2}]^2+6[\accentset{(\alpha)}{\Pi}^{4}]\notag\\
    &=[\Pi]^4-6[\Pi]^2[\Pi^{2}]+8[\Pi][\Pi^{3}]+3[\Pi^{2}]^2+6[\Pi^{4}]\notag\\
    & \quad-8\alpha(\Box\pi)^3\Box^2\pi\notag\\
    & \quad+24\alpha\big(\partial_{\mu}\partial_{\alpha}\pi\Box\partial^{\mu}\partial^{\alpha}\pi(\Box\pi)^2+\Box^2\pi\Box\pi(\partial^{\mu}\partial^{\alpha}\pi)^2\big)\notag\\
    & \quad -8\alpha\big(6\Box\pi\Box^2\pi(\partial_{\mu}\partial^{\alpha}\pi)^2+2\Box^2\pi(\partial_{\mu}\partial^{\alpha}\pi)^3\big)\notag\\
    & \quad -24\alpha(\partial_{\mu}\partial^{\alpha}\pi)^3\Box\partial^{\mu}\partial_{\alpha}\pi\notag\\
    & \quad -48\alpha(\partial_{\mu}\partial^{\alpha}\pi)^3\Box\partial^{\mu}\partial_{\alpha}\pi~.
\end{align}
\section{Discussion and Conclusion}
\par\noindent
The analysis presented in this work demonstrates that a coherent picture emerges when GUP-inspired short-distance modifications are applied to the dynamics of a massless spin-2 field. At the level of quadratic fluctuations, the GUP deformation reorganizes the FP action into a higher-derivative structure whose content is indistinguishable from the spin-2 sector of Stelle's curvature-squared theory. What is nontrivial is that the appearance of the massive spin-2 excitation is not imposed by hand but arises directly from the deformation of the momentum operator implied by the GUP. In this sense, the GUP acts as a UV regulator \cite{Kempf:1996nk}, whose footprint on the spin-2 propagator reproduces, in a controlled way, the same analytic structure that is normally attributed to curvature-squared corrections in effective gravity.
\par\noindent
The auxiliary-field formulation makes this UV/IR interplay explicit. In GUP language, the deformation introduces an additional short-distance scale which manifests as a nonlocal modification of the differential operator. In the higher-derivative description, the same modification appears as an additional mode propagating at intermediate distances with a wrong-sign kinetic term. Under LW or PT-symmetric quantization, this mode is removed from the asymptotic Hilbert space and contributes only virtually, thereby preserving perturbative unitarity. In effect, the GUP deformation translates a microscopic length scale -normally associated with Planck-scale quantum geometry- into an IR modification of the spin-2 propagator through the emergence of this LW partner. Therefore, the theory  provides a clean example of how Planckian physics may leave detectable imprints at scales far below the UV cutoff, not through new propagating states, but through modified analytic structure.
\par\noindent
A second facet of this UV/IR relation appears in the nonlinear sector. The decoupling-limit interactions of dRGT massive gravity are highly constrained: the cancellation of the Boulware-Deser ghost relies on a delicate balance among the Galileon terms. Remarkably, the GUP deformation considered here preserves this structure at $\mathcal{O}(\alpha)$ with all additional pieces collapsing into total derivatives. This means that the same UV mechanism that introduces a short-distance improvement in the spin-2 propagator does not spoil the IR consistency of the nonlinear massive theory. Instead, the GUP leaves the helicity-0 sector intact to leading order, suggesting that certain classes of UV corrections may remain compatible with ghost-free massive gravity. This compatibility is nontrivial; a generic higher-derivative deformation would reintroduce the Boulware-Deser mode, whereas the GUP, acting through the deformed kinetic term, generates only harmless boundary terms. We speculate the GUP regime would show the same behavior when applied to the massive dual spin-2 fields \cite{Danehkar:2021obr,Curtright:2019wxg,Alshal:2019hpk}.
\par\noindent
Despite we dropped the $G,\hbar,c$ constants in the equations, it is worth noting that the GUP-induced structure obtained here parallels, in a conceptually evident way, the mechanism proposed by Minkowski for the \emph{spontaneous} origin of Newton's constant~\cite{Minkowski:1977aj}. In Minkowski's framework, a nonminimal coupling of scalar fields to curvature, leads after spontaneous symmetry breaking to an emergent Newton constant so that the Einstein--Hilbert term arises as an IR residue of UV scalar dynamics. No fundamental gravitational coupling is inserted in the microscopic theory; the Planck scale appears only as the vacuum value of the underlying field configuration.
\par\noindent
The GUP deformation realizes an analogous IR emergence, though through a different microscopic mechanism. The modified momentum operator generates the higher-derivative operator, whose low-momentum limit reproduces Einstein gravity while simultaneously introducing a massive Lee-Wick partner with mass. In this sense, the effective Newton constant is again determined by a UV parameter - here the GUP parameter scales as  $\alpha\sim \ell^{-2}_{\text{min}}$ - rather than by a fundamental Einstein-Hilbert term.
\par\noindent
Both constructions therefore embody the same structural principle; \emph{Newton's constant is an emergent IR scale encoded in UV physics}. Minkowski achieves this through spontaneous symmetry breaking in a scalar sector, while the present work achieves it through a minimal-length deformation of phase space whose higher--derivative imprint mimics the quadratic sector of Stelle's gravity. Moreover, the softness of the UV modification -essential in Minkowski's argument- is mirrored here by the fact that GUP corrections reduce to total derivatives within the Galileon interactions of the dRGT helicity-0 mode, preserving second-order equations of motion and avoiding the Boulware-Deser ghost. 
Thus, the GUP-LW picture developed here may be understood as a non-scalar but structurally analogous realization of Minkowski's emergent gravity paradigm.
\par\noindent
Taken together, these results point toward a deeper structural relation between UV-complete descriptions of geometry and their IR effective field theories. The GUP's minimal-length scale propagates downward into the IR as a LW-type regulator for the spin-2, and a restricted, symmetry-compatible deformation of Galileon interactions. This dual behavior indicates that Planck-suppressed operators need not disrupt IR consistency if they enter through a deformation of the canonical phase-space algebra rather than through arbitrary additions to the Lagrangian. The fact that the same deformation reproduces both the Stelle operator and leaves dRGT's ghost-free structure intact suggests a possible guiding principle for constructing viable UV completions of massive gravity: the underlying short-distance theory may need to be encoded through algebraic (GUP-like) rather than purely Lagrangian deformations.
\par\noindent
Future work should investigate whether this GUP--LW correspondence persists beyond quadratic order, whether the nonperturbative contour prescriptions can be embedded in a covariant quantization of the metric, and whether the GUP-generated scale leaves observable signatures in cosmological or astrophysical settings. Nonetheless, the present results illustrate that UV modifications driven by a generalized uncertainty principle can generate IR-complete and unitary massive spin-2 theories, providing a bridge between minimal-length quantum geometry and low-energy gravitational phenomenology.
%
%
%
%
%

\end{arabicfootnotes}
\end{document}